\documentclass[prl,twocolumn,preprintnumbers,amsmath,amssymb,showpacs,
nofootinbib,floatfix]{revtex4}

\usepackage{graphicx, bm}

\usepackage[normalem]{ulem}
\usepackage{comment}

\makeatletter
\def\graphicscale{\twocolumn@sw{0.33}{0.4}}
\makeatother

\usepackage{ifthen}
\usepackage{hyperref}
\newboolean{ARXIV}
\setboolean{ARXIV}{true}
\newcommand{\arxiv}[1]{\ifthenelse{\boolean{ARXIV}}{ [\href{http://arxiv.org/abs/#1}{#1}]}{}}
\newcommand{\arxivonly}[1]{\ifthenelse{\boolean{ARXIV}}{#1}{}}

\begin{document}

\title{Change of $\theta$ dependence in 4D SU($N$) gauge theories 
across the deconfinement transition}

\author{Claudio Bonati,$^{1}$ Massimo D'Elia,$^1$ 
Haralambos Panagopoulos,$^2$ and Ettore Vicari$^1$ }

\affiliation{$^1$ Dip. Fisica dell'Universit\`a di Pisa and INFN,
Largo Pontecorvo 3, I-56127 Pisa, Italy}

\affiliation{$^2$ Department of Physics, University of Cyprus, Lefkosia,
CY-1678, Cyprus }

\begin{abstract}
We study the dependence of 4D SU($N$) gauge theories on
the topological $\theta$ term at finite temperature $T$.
We exploit the lattice formulation of the theory,
presenting numerical results for the expansion of
the free energy up to $O(\theta^6)$, for $N=3$ and $N=6$.
Our analysis shows that the $\theta$ dependence drastically
changes across the deconfinement transition: the low-$T$ phase
is characterized by a large-$N$ scaling with $\theta/N$ as relevant
variable, while in the high-$T$ phase the scaling variable is
just $\theta$ and the free energy is essentially determined
by the instanton-gas approximation.
\end{abstract}


\maketitle


Important physical issues of strong interactions are related to the
nontrivial dependence of 4D SU($N$) gauge theories on the topological
parameter $\theta$, which appears in the Euclidean Lagrangian as
\begin{equation}
{\cal L}_\theta  = \frac{1}{4} F_{\mu\nu}^a(x)F_{\mu\nu}^a(x)
- i \theta \frac{g^2}{64\pi^2} \epsilon_{\mu\nu\rho\sigma}
F_{\mu\nu}^a(x) F_{\rho\sigma}^a(x),
\label{lagrangian}
\end{equation}
where 
\begin{equation}
q(x)=\frac{g^2}{64\pi^2} 
\epsilon_{\mu\nu\rho\sigma} F_{\mu\nu}^a(x) F_{\rho\sigma}^a(x)
\label{topchden}
\end{equation}
is the topological charge density.  The presence of a nonzero $\theta$
would break both parity and time reversal and the experimental upper
bound on it is very small, $|\theta| < 10^{-9}$~\cite{Baker}.
Nevertheless, the issue of $\theta$ dependence is interesting and
relevant to hadron phenomenology, an example being the so-called
U($1$)$_A$ problem. In the framework of the large-$N$
expansion~\cite{Hooft-74,Witten-79,Veneziano-79}, the nontrivial
$\theta$ dependence provides an explanation to the fact that the
U($1$)$_A$ symmetry of the classical theory is not realized in the
hadron spectrum (see, e.g., Refs.~\cite{VP-09,LP-12} for recent
reviews).

In this paper we investigate the topological properties and the
$\theta$ dependence of 4D SU($N$) gauge theories at finite temperature
$T$, in particular across the deconfining temperature $T_c$. Such
properties are known to be relevant to the thermodynamic behavior of
hadronic matter.  For example, the effective restoration of the
U($1$)$_A$ symmetry in strong interactions at finite $T$, and in
particular around the chiral transition, is relevant to the nature of
the transition itself~\cite{VP-09,piwi,BPV-05}.

As we shall better discuss in the following, one expects, on general
grounds, a crossover between a low-$T$ and a high-$T$ regime for
$\theta$ dependence, characterized by different large-$N$ scalings.
In particular, the high-$T$ regime should be describable by a
semiclassical instanton gas picture, which instead fails in the
low-$T$ regime, where $\theta/N$ turns out to be the relevant
large-$N$ scaling variable.  In 4D SU($N$) gauge theories the
deconfining transition is first order for $N \geq 3$ and gets stronger
as $N$ increases. It is therefore reasonable to conjecture that the
$\theta$ dependence may sharply change right around $T_c$, where one
may expect a singular behavior, such as a discontinuity.

The purpose of our study is to investigate such a scenario
numerically, presenting results for $N=3$ and $N=6$ to check the $N$
dependence around the deconfinement transition.  The lowest order
$O(\theta^2)$ contribution to the free energy, involving just the
topological susceptibility, has been already investigated around the
deconfinement transition~\cite{susc_ft,GHS-02,DPV-04,LTW-05}. However,
as we shall discuss in detail later on, the study of higher order
terms provides a more stringent and definite signature for the change
of the $\theta$ dependence between the two phases. Such a study is the
main subject of our investigation.

The finite-$T$ behavior is specified by the free energy
\begin{equation}
F(\theta,T) = 
-\frac{1}{\cal V} \ln \int [dA] 
\exp \left(  - \int_0^{1/T} dt \int d^3 x\, {\cal L}_\theta \right),
\label{vftheta}
\end{equation}
where $T$ is the temperature, ${\cal V}=V/T$ is the
Euclidean space-time volume, and the gluon field satisfies
$A_\mu(1/T,{\bf x}) = A_\mu(0,{\bf x})$. The $\theta$ dependence can be
parameterized as 
\begin{eqnarray}
{\cal F}(\theta,T)\equiv F(\theta,T)-F(0,T)={1\over 2} \chi(T)
\theta^2 s(\theta,T),\label{ftheta}
\end{eqnarray}
where $\chi(T)$ is the topological susceptibility at $\theta=0$,
\begin{eqnarray}
&&\chi = \int d^4 x \langle q(x)q(0) \rangle_{\theta=0} 
= {\langle Q^2 \rangle_{\theta=0} \over {\cal V}},\label{chidef}
\end{eqnarray}
and $s(\theta,T)$ is a dimensionless even function of $\theta$ such
that $s(0,T)=1$.  Assuming analyticity at $\theta=0$, $s(\theta,T)$
can be expanded as
\begin{eqnarray}
s(\theta,T) = 1 + b_2(T) \theta^2 + b_4(T) \theta^4 + \cdots,
\label{stheta}
\end{eqnarray}
where only even powers of $\theta$ appear.  

At zero temperature, where the free energy coincides with the
ground-state energy, large-$N$ scaling
arguments~\cite{Hooft-74,Witten-98} applied to the Lagrangian
(\ref{lagrangian}) indicate that the relevant scaling
variable~\cite{footnoteln} is $\bar\theta\equiv {\theta/N}$, i.e.
\begin{equation} 
{\cal F}(\theta) \approx  N^2 {\cal G}(\bar{\theta})
\label{fthetabarln} 
\end{equation}
as $N\to \infty$.  Comparing with Eq.~(\ref{ftheta}), this implies the
large-$N$ behavior
\begin{equation}
\chi = \chi_\infty + O(N^{-2}), \quad  b_{2j}=O(N^{-2j}). 
\label{lnasyt0}
\end{equation}
We recall that a nonzero value of $\chi_{\infty}$ is essential to
provide an explanation to the U($1$)$_A$ problem in the large-$N$
limit~\cite{Witten-79,Veneziano-79}.  The apparent incompatibility of
Eq.~\eqref{fthetabarln} with the periodic nature of the topological
$\theta$ term may be solved by a nonanalytic multibranched $\theta$
dependence of the ground-state energy~\cite{Witten-80,Witten-98},
$F(\theta) = N^2 \, {\rm min}_k\, H\left( {\theta+2\pi k\over
N}\right)$.  The large-$N$ scaling (\ref{fthetabarln}) of the $\theta$
dependence is fully supported by numerical computations exploiting the
nonperturbative Wilson lattice formulation of the 4D SU($N$) gauge
theory at $T=0$, see, e.g., the results reported in Table~\ref{t0res}
for $N=3,4,6$ (see also Refs.~\cite{VP-09,LP-12} for recent reviews).
This scenario is expected to remain stable against sufficiently low
temperatures.

\begin{table}
\footnotesize
\caption{Summary of known $T=0$ results for the ratio $\chi/\sigma^2$
(where $\sigma$ is the string tension) and the first coefficients
$b_{2j}$ for $N=3,4,6$~\cite{n6t0}. The extrapolation of the results
for $\chi/\sigma^2$, using the simple Ansatz $a+b/N^2$, gives
$\chi/\sigma^2=0.022(2)$ for $N\to\infty$.  For more complete reviews
of results see Refs.~\cite{VP-09,LP-12}.  }
\label{t0res}
\begin{ruledtabular}
\begin{tabular}{clll}
\multicolumn{1}{c}{$N$}&
\multicolumn{1}{c}{$\chi/\sigma^2$}&
\multicolumn{1}{c}{$b_2$}&
\multicolumn{1}{c}{$b_4$}\\
\colrule
3 & 0.028(2) \cite{VP-09} 
& $-$0.026(3) \cite{PV-11,Delia-03,GPT-07} &  0.000(1) \cite{PV-11} \\
4 & 0.0257(10) \cite{DPV-02} & $-$0.013(7) \cite{DPV-02} & \\
6 & 0.0236(10) \cite{DPV-02} & $-$0.008(4) & 0.001(3) 
\end{tabular}
\end{ruledtabular}
\end{table}

The large-$N$ scaling (\ref{fthetabarln}) is not realized by the
dilute instanton gas approximation. Indeed, at zero
temperature, instanton calculations fail due to the fact that large
instantons do not get suppressed. On the other hand, temperature acts
as an infrared regulator, so that the instanton-gas partition function
is expected to provide an effective approximation of finite-$T$
SU($N$) gauge theories at high temperature~\cite{GPY-81}, high enough
to make the overlap between instantons negligible.  The corresponding
$\theta$ dependence is~\cite{GPY-81,CDG-78}
\begin{eqnarray}
&&{\cal F}(\theta,T) \approx \chi(T) \left( 1 - \cos\theta\right), 
\label{thdepht}\\
&&
\chi(T) \approx  T^4 \exp[-8\pi^2/g^2(T)] \sim  T^{-\frac{11}{3} N + 4}, 
\label{chitasy}
\end{eqnarray}
using $8 \pi^2/g^2(T) \approx (11/3) N \ln (T/\Lambda)+O(\ln\ln
T/\ln^2 T)$.  Therefore, the high-$T$ $\theta$ dependence
substantially differs from that at $T=0$\,: the relevant variable
for the instanton gas is just $\theta$, and not $\theta/N$.  The
instanton-gas approximation also shows that $\chi(T)$, and therefore
the instanton density, gets exponentially suppressed in the large-$N$
regime, thus suggesting a rapid decrease of the topological activity
with increasing $N$ at high $T$. Since the instanton density gets
rapidly suppressed in the large-$N$ limit, making the probability of
instanton overlap negligible, the range of validity of the
instanton-gas approximation is expected to rapidly extend toward
smaller and smaller temperatures with increasing $N$.

The low-$T$ and high-$T$ phases are separated by a first-order
deconfinement transition which becomes stronger with increasing $N$
\cite{LTW-04}, with $T_c$ converging to a finite large-$N$
limit~\cite{LRR-12}.  This suggests the following scenario: the
crossover between the low-$T$ large-$N$ scaling $\theta$ dependence
and the high-$T$ instanton-gas $\theta$ dependence, respectively given
by Eqs. (\ref{fthetabarln}) and (\ref{thdepht}), occurs around the
deconfinement transition, and becomes sharper and sharper with
increasing $N$. See, e.g., Refs.~\cite{KPT-98,DPV-04,BL-07,PZ-08} for further
discussions of this scenario.

It is important, at this point, to stress the following: even if the
instanton-gas prediction, Eq.~(\ref{thdepht}), receives significant
corrections as one approaches $T_c$ from above, one can still
conjecture that the phase transition sharply delimits two regimes with
a different large-$N$ scaling behavior, i.e. that the free energy is a
function of $\theta/N$ in the confined phase and a function of
$\theta$ in the deconfined phase.  This conjecture is of course more
general than the instanton gas picture itself.

The finite-$T$ lattice investigations of the large-$N$ behavior of
$\chi(T)$~\cite{DPV-04,LTW-05} indicate a nonvanishing large-$N$ limit
for $T<T_c$, remaining substantially unchanged in the low-$T$ phase,
from $T=0$ up to $T_c$.  Across $T_c$ a sharp change is observed, and
$\chi(T)$ appears largely suppressed in the high-$T$ phase $T>T_c$, in
qualitative agreement with a high-$T$ scenario based on the
instanton-gas approximation.

However, to achieve a more stringent check of the actual scenario
realized in 4D SU($N$) gauge theories, we consider the higher-order
terms of the expansion \eqref{stheta}, which provide further
significant information on the $\theta$ dependence.  Indeed, the
expansion coefficients $b_{2j}$ are expected to scale like $N^{-2j}$
if the free energy is a function of $\theta/N$ and to be
$N$-independent in the instanton-gas approximation.  The finite-$T$
behavior of such coefficients has never been studied numerically until
now and is the subject of our investigation.  In particular, the
simple $\theta$ dependence of Eq.~(\ref{thdepht}) may be observed at
much smaller $T$ above $T_c$ with respect to the asymptotic one-loop
behavior (\ref{chitasy}) of $\chi(T)$ which is subject to logarithmic
corrections.

In particular, we aim at clarifying: {\em i)} whether a sharp change
in the large-$N$ scaling of $b_{2j}$ is observed across $T_c$,
signalling a change from a $\theta/N$ to a $\theta$ dependence of the
free energy; {\em ii)} how rapidly the values of $b_{2j}$ above $T_c$
converge to the instanton gas prediction, Eq.~\eqref{thdepht}, i.e.
\begin{equation}
b_{2j} = (-1)^j {2\over(2j+2)!},  \quad j=1,2,...,
\label{b24ig}
\end{equation}
for the expansion (\ref{stheta}). These predictions should be compared
to the $T=0$ estimates summarized in Table~\ref{t0res}.  It has to be
stressed that the results \eqref{b24ig} depend just on the form
\eqref{thdepht} of the free energy and are independent of the
renormalized coupling costant $g(T)$. 

Due to the nonperturbative nature of the physics of $\theta$
dependence, quantitative assessments of this issue have largely
focused on the lattice formulation of the theory, using Monte Carlo
(MC) simulations.  However, the complex character of the $\theta$ term
in the Euclidean QCD Lagrangian prohibits a direct MC simulation at
$\theta\ne 0$.  Information on the $\theta$ dependence of physically
relevant quantities, such as the ground state energy and the spectrum,
can be obtained by computing the coefficients of the corresponding
expansion around $\theta = 0$.

We mention that issues related to $\theta$ dependence, particularly in
the large-$N$ limit, can also be addressed by other approaches, such
as AdS/CFT correspondence applied to nonsupersymmetric and
nonconformal theories, see
e.g. Refs.~\cite{AGMOO-00,Witten-98,kiritsis-etal,PZ-08}, and
semiclassical approximation of compactified gauge
theories~\cite{ZT-12, unsal}.

In order to check the change of $\theta$ dependence across the
deconfinement transition, we numerically compute the topological
susceptibility and the first few coefficients of the expansion
(\ref{stheta}) above $T_c$, for $N=3$ and $N=6$ to check the $N$
dependence.  For this purpose we exploit the lattice Wilson
formulation of SU($N$) gauge theories
\begin{equation}
S_L = - {\beta\over N} \sum_{x,\mu>\nu} {\rm Re} {\rm Tr}\,\Pi_{\mu\nu}(x),
\label{lattaction}
\end{equation}
where $\Pi_{\mu\nu}$ is the standard plaquette operator \cite{Wilson}.
The coefficients of the expansion around $\theta=0$ can be determined
from appropriate zero-momentum correlation functions of the
topological charge density at $\theta=0$. These are related to the
moments of the $\theta=0$ probability distribution $P(Q)$ of the
topological charge $Q$ and parameterize the deviations of $P(Q)$ from
a simple Gaussian behavior.  Indeed~\cite{errata},
\begin{eqnarray}
&&\chi_l = \frac{\langle Q^2 \rangle}{L_tL_s^3}, \label{chil}\qquad 
b_2 = -\, { \langle Q^4 \rangle - 3  \langle Q^2 \rangle^2  \over 
12 \langle Q^2 \rangle } , 
\label{b2chi4} \\
&&b_4 =  {\langle Q^6 \rangle  -  15 \langle Q^2 \rangle \langle Q^4 \rangle  +
30 \langle Q^2 \rangle^3 
\over 360 \langle Q^2 \rangle} ,
\label{b4chi6} 
\end{eqnarray}
where $\chi_l$ is the the lattice topological susceptibility
($\chi_l\approx a^4 \chi$; $a$ is the lattice spacing).  The
correlation functions involving multiple zero-momentum insertions of
the topological charge density can be defined in a nonambiguous,
regularization independent way~\cite{Luscher-04}, and therefore the
expansion coefficients $b_{2i}$ are well defined renormalization-group
invariant quantities.  This implies that they approach their continuum
limit with $O(a^2)$ corrections.

We evaluate the above quantities in MC simulations for several values
of the coupling $\beta$ on asymmetric $L_t\times L_s^3$
lattices~\cite{footnoteup}.  Accurate estimates of $b_{2j}$ require
huge statistics, because of the large cancellations when evaluated
from the expectation values of powers of $Q$, as in
Eq.~(\ref{b2chi4}).  Therefore, we have to consider a relatively fast
method to estimate the topological charge $Q$ of a lattice
configuration.  We choose the cooling method, and in particular the
implementation outlined in Ref.~\cite{DPV-02}.  The stability of the
results under cooling is carefully checked; we take our data after 15
cooling steps, but differences with the results after 10 and 20
cooling steps remain always within the errors reported~\cite{Ntsize}\arxivonly{ (for
some examples see the appendix)}.
Moreover, the
stability substantially improves with increasing $N$, as already noted
in the literature, also by detailed comparisons with the more rigorous
overlap method (which is much more demanding numerically), see, e.g.,
Ref.~\cite{VP-09}.

A summary of the results for $N=3$ and $N=6$ is presented in
Tables~\ref{ftN3} and \ref{ftN6} respectively~\cite{footnotestat}.
The aspect ratio $L_s/L_t$ in our MC simulations is sufficiently large
to give rise to infinite-volume results for the observables considered
within the statistical errors, as shown by the comparison of results
for different values of $L_s$.  For the case of both SU($3$) and
SU($6$) we check the continuum limit by comparing the results obtained
by using two lattices of different temporal extent at the largest
value of $T$, see Tab.~\ref{ftN3}-\ref{ftN6}.

\begin{table}
\footnotesize
\caption{$N=3$ results. We report the value of lattice coupling
$\beta$, the temporal ($L_t$) and spatial ($L_s$) size of the lattice,
the reduced temperature $t\equiv T/T_c-1$~\cite{footnotetcres},
$\chi_l$, $\chi/T_c^4$ and the first two coefficients $b_2$ and $b_4$ of the
expansion (\ref{stheta}).  }
\label{ftN3}
\begin{ruledtabular}
\begin{tabular}{lclllll}
\multicolumn{1}{c}{$\beta$}& \multicolumn{1}{c}{$L_t, L_s$}&
\multicolumn{1}{c}{$t$}& \multicolumn{1}{c}{$10^5\chi_l$}&
\multicolumn{1}{c}{$10\,\chi/T_c^4$}&
\multicolumn{1}{c}{$-12\,b_2$}& \multicolumn{1}{c}{$360\,b_4$}\\ 
\colrule 
6.173 & 10, 40 & $-$0.053(3)           & 2.292(7)  & 1.84(2)   & 0.37(12) & $-4(11)$ \\ 

6.241 & 10, 40 & \phantom{$-$}0.045(3) &  0.645(3) & 0.77(1)   & 1.27(7) & \phantom{$-$}0.7(1.8) \\ 

6.273 & 10, 40 & \phantom{$-$}0.095(4) &  0.375(3) & 0.54(1)   & 1.15(7) & \phantom{$-$}1.4(1.4) \\ 

6.305 & 10, 40 & \phantom{$-$}0.145(6) &  0.232(2) & 0.40(1)   & 1.02(5) & \phantom{$-$}3.6(7.2) \\ 

6.305 & 10, 30 & \phantom{$-$}0.145(6) &  0.233(3) & 0.40(1)   & 1.10(7) & \phantom{$-$}2.9(1.4) \\ 

6.437 & 12, 48 & \phantom{$-$}0.147(13) & 0.103(3) & 0.37(2)   & 1.07(14) & $-$1.1(1.4)
\end{tabular}
\end{ruledtabular}
\end{table}

\begin{table}
\footnotesize
\caption{$N=6$ results. We report the same quantities as in
Table~\ref{ftN3}~\cite{footnotetcres}. }
\label{ftN6}
\begin{ruledtabular}
\begin{tabular}{lclllll}
\multicolumn{1}{c}{$\beta$}&
\multicolumn{1}{c}{$L_t,L_s$}&
\multicolumn{1}{c}{$t$}&
\multicolumn{1}{c}{$10^5\chi_l$}&
\multicolumn{1}{c}{$10^3\chi/T_c^4$}&
\multicolumn{1}{c}{$-12\,b_2$}&
\multicolumn{1}{c}{$360\,b_4$}\\
\colrule
24.797 & 6, 24 & $-$0.032(10)             & 17.14(16)  & 195(8)   & 0.07(34) & $-$14(18) \\

24.912 & 6, 24 & \phantom{$-$}0.045(14)   & 0.622(13)  & 9.6(5)   & 1.15(8)  & 2.2(1.8) \\

24.912 &   6, 20 & \phantom{$-$}0.045(14) & 0.631(16)  & 9.8(6)   & 1.17(8)  & 2.2(0.7) \\

25.056 &  6, 24 &  \phantom{$-$}0.089(8)  & 0.132(3)   &  2.41(9) & 1.02(4)  & 1.0(2) \\

24.768 &  5, 20 &  \phantom{$-$}0.141(7)  & 0.121(3)   &  1.28(4) & 1.02(2)  &  1.0(1) \\

25.200 &  6, 24 &  \phantom{$-$}0.160(8)  & 0.0316(12) &  0.74(3) & 1.02(4)  & 1.1(1) 
\end{tabular}
\end{ruledtabular}
\end{table}

\begin{figure}[tbp]
\includegraphics*[scale=\graphicscale]{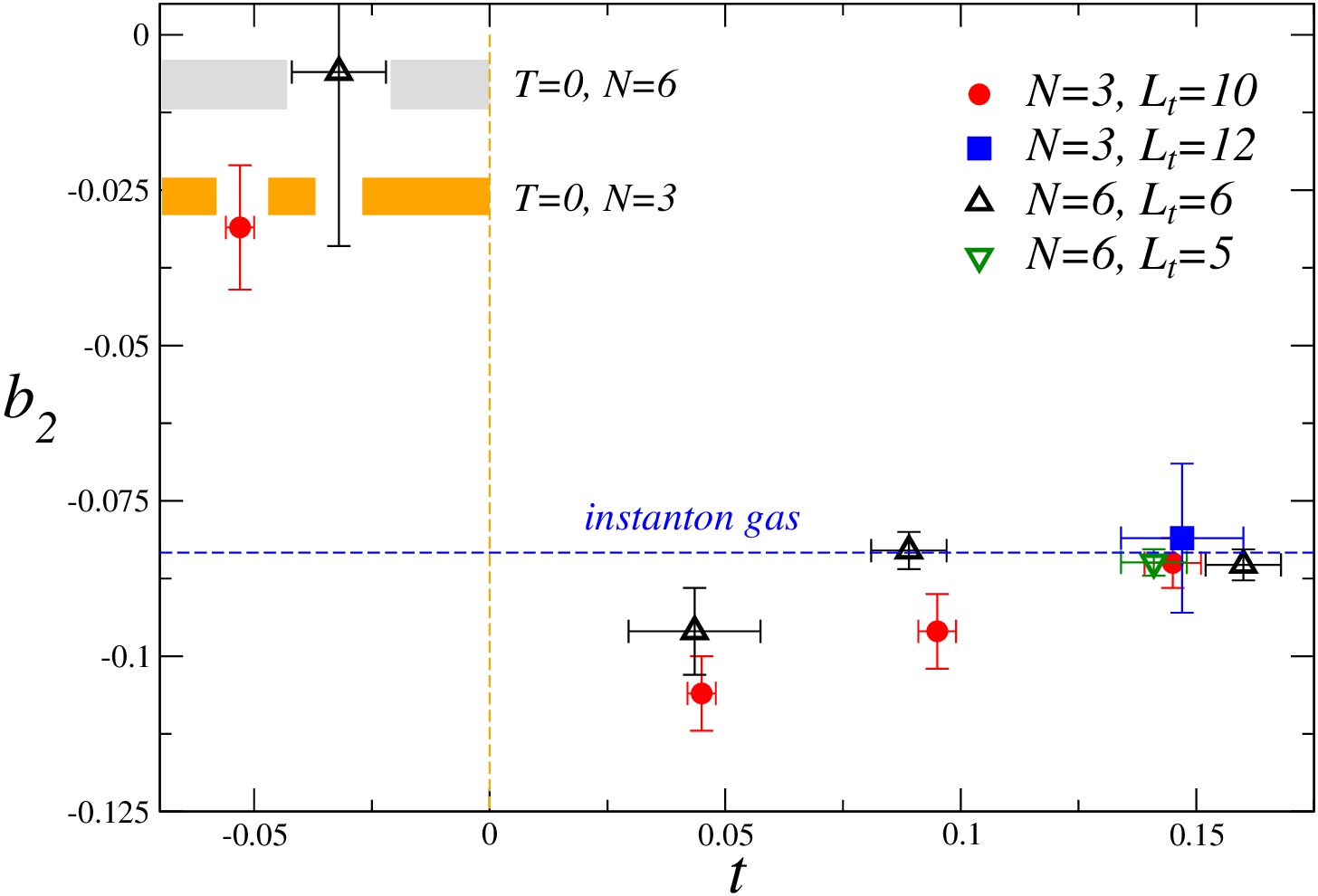}
\caption{(Color online) Finite-$T$ results of $b_2$ for $N=3$ and
$N=6$, versus the reduced temperature $t\equiv T/T_c-1$.  The shadowed
regions indicate the $T=0$ estimates of $b_2$ for $N=3$ and $N=6$, see
Table~\ref{t0res}.  }
\label{b2fig}
\end{figure}

The MC results clearly show a change of regime in the $\theta$
dependence, from a low-$T$ phase where the susceptibility and the
coefficients of the $\theta$ expansion vary very little, to a high-$T$
phase where the coefficients $b_{2j}$ approach the instanton-gas
predictions.  Fig.~\ref{b2fig} shows the data for $b_2$.  In the
high-$T$ phase they are definitely not consistent with the scaling
\eqref{lnasyt0}, which would imply a factor of four in $b_2$, in going
from $N=3$ to $N=6$.  On the other hand, in the low-$T$ phase $b_2$
does not significantly differ from the $T=0$ value.  This is
consistent with the behaviour of $\chi_l$ at $N=3$, for which we
obtain: $\chi_l(T=0.95 T_c)/\chi_l(T=0) = 0.98(1)$ (at
$\beta=6.173$). A similar behaviour is observed for $N=6$:
$\chi_l(T=0.97T_c)/\chi_l(T=0) = 1.00(2)$ (at $\beta=24.797$).

Although our MC results in the
high-$T$ phase are obtained for relatively small reduced temperatures
$t\equiv T/T_c-1$, i.e. $t < 0.2$, the data for $b_2$ show a clear and
rapid approach to the value $b_2=-1/12$ of the instanton gas model for
both $N=3$ and $N=6$, with significant deviations visible only for
$t\lesssim 0.1$.  The high-$T$ values of $b_2$ substantially differ
from those of the low-$T$ phase, and in particular from those at $T=0$
reported in Table~\ref{t0res}.  Also the estimates of $b_4$ are
consistent with the small value $b_4=1/360$.

Our data confirm that $\chi$ rapidly decays with increasing $t$ in
both $N=3,6$ cases. In particular for $N=6$ we obtain $\chi_l(T=1.09
T_c)/\chi_l(T=0) = 0.0136(4)$ (at $\beta=25.056$).  This suppression
is in qualitative agreement with the one-loop instanton-gas result
(\ref{chitasy}), but larger temperatures are required for a reliable
quantitative comparison, essentially because of the logarithmic
corrections to Eq.~(\ref{chitasy}).  The sharp behavior of the
$\theta$ dependence at the phase transition suggests that $T_c$ is
actually a function of $\theta/N$, as put forward in
Ref.~\cite{DN-12}.

We have tried to understand the deviations for $b_2$, visible at $t
\lesssim 0.1$, by taking into account corrections to the instanton-gas
formula (\ref{thdepht}) through a virial-like expansion: the
asymptotic formula is corrected by a term proportional to the square
of the instanton density.  For example, we may write
\begin{equation}
{\cal F}(\theta,T) \approx \chi (1-\cos\theta) + 
\chi^2 \kappa(\theta) + O(\chi^3),
\label{fthcorr}
\end{equation}
where we use the fact that $\chi(T)$ is proportional 
to the instanton density, and
$\kappa(\theta)$ can be parametrized as
$\kappa(\theta)=\sum_{k=2} c_{2k} \sin(\theta/2)^{2k}$. 
The above formula gives
\begin{equation}
b_2 = -{1\over 12} + {1\over 8}\, c_4 \frac{\chi}{T_c^4} + O\left(\frac{\chi^2}{T_c^8}\right).
\label{b2corr}
\end{equation}
If $\chi$ gets rapidly suppressed in the high-$T$ phase, as suggested
by Eq.~\eqref{chitasy} and confirmed by the MC results,
Eq.~\eqref{b2corr} would imply a rapid approach to the asymptotic
value of the perfect instanton gas, as shown by the data, see
Fig.~\ref{b2fig}.  Assuming $c_4$ weakly dependent on $N$, Eq.~\eqref{b2corr}
predicts an exponentially faster convergence with increasing $N$, 
as also supported by the data.
Moreover, a hard-core approximation of the
instanton interactions~\cite{CDG-78} gives rise to a negative
correction, i.e.  $c_4<0$, explaining the approach from below to the
perfect instanton-gas value $b_2=-1/12$.

In conclusion, our numerical analysis provides strong evidence that
the $\theta$ dependence of 4D SU($N$) gauge theory experiences a
drastic change across the deconfinement transition, from a low-$T$
phase characterized by a large-$N$ scaling with $\theta/N$ as relevant
variable, to a high-$T$ phase where this scaling is lost and the free
energy is essentially determined by the instanton-gas approximation,
which implies an analytic and periodic $\theta$ dependence. The
corresponding crossover around the transition becomes sharper with
increasing $N$ (see Fig.~\ref{b2fig}), suggesting that the perfect instanton-gas regime sets
in just above $T_c$ at large $N$, while $\chi(T)$ gets drastically
suppressed. A virial-like expansion suggests that the approach is   
exponential in $N$; this issue deserves further investigation.

It is interesting to remark that hints for an early onset of 
an instanton gas regime above the chiral/deconfinement
transition have been provided by recent MC simulations of full QCD,
by looking at the behavior of the relevant susceptibilities~\cite{ua1refs}.

{\em Acknowledgements.} A large part of the MC simulations was
performed at the INFN Pisa GRID DATA center, using also the cluster
CSN4, for a total amount of about $600$ years of CPU time.  HP would
like to thank the Research Promotion Foundation of Cyprus for support,
and INFN, Sezione di Pisa, for the kind hospitality.  We thank
Francesco Bigazzi and Ariel Zhitnitsky for enlightening discussions.
CB and MD thank the Galileo Galilei Institute for Theoretical Physics
for the hospitality offered during the workshop ''New Frontiers in
Lattice Gauge Theories".

\arxivonly{
\section{Appendix:
stability analysis of the topological measurements}

\begin{figure}[htbp]
\includegraphics*[scale=\graphicscale]{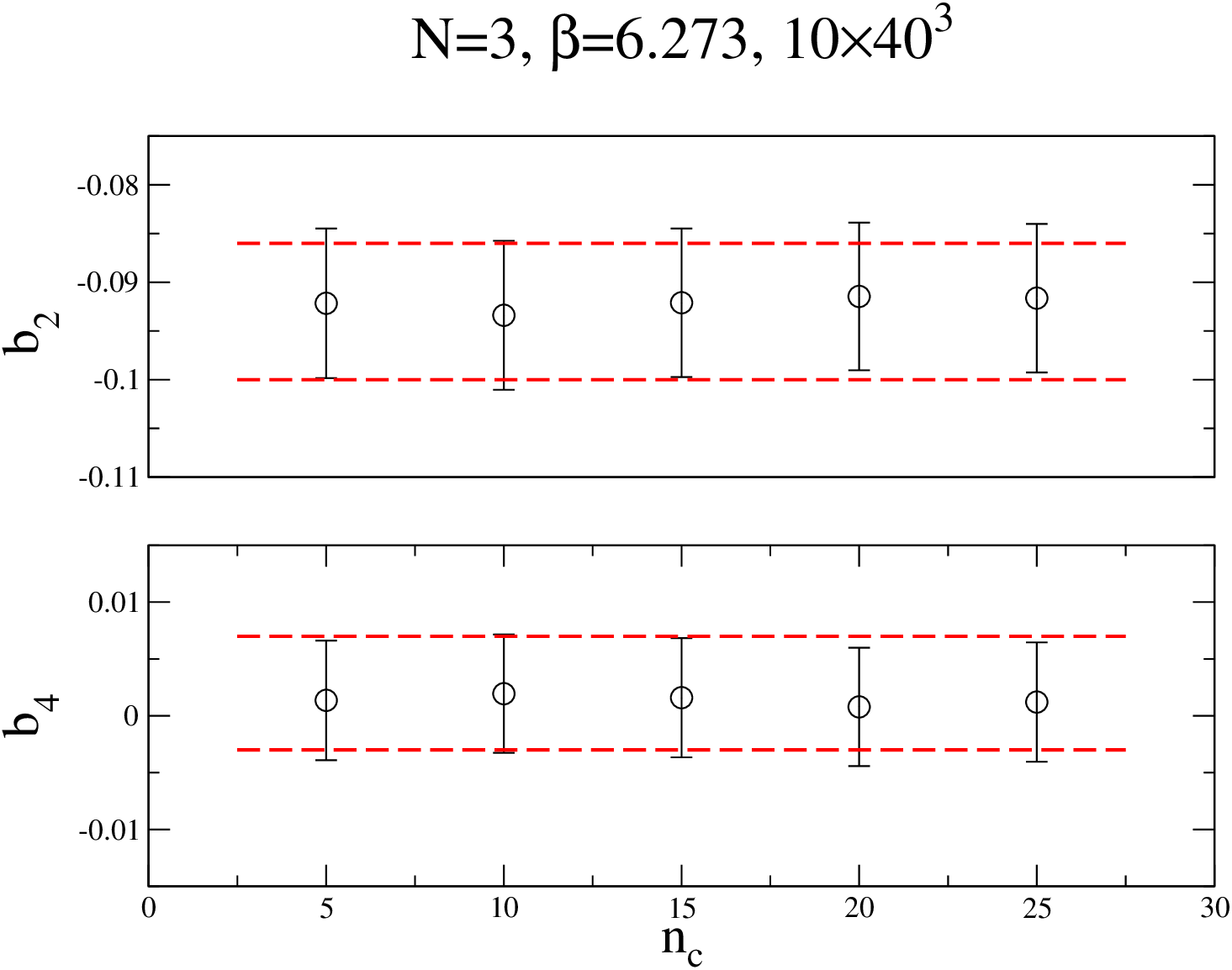}
\caption{Results for SU($3$).}
\label{test_3}
\end{figure}

In Fig. \eqref{test_3} and \eqref{test_6} the stability of the results
presented in the paper for $b_2$ and $b_4$ under cooling is shown:
$n_c$ is the number of cooling steps and $b_2$ and $b_4$ are estimated
by means of \eqref{b4chi6}, using for the determination of topological
charge $Q$ the prescription of \cite{DPV-02}.  Figures refer to the
case $t\approx 0.1$, but all the simulations present similar
behaviour.

\begin{figure}[htbp]
\includegraphics*[scale=\graphicscale]{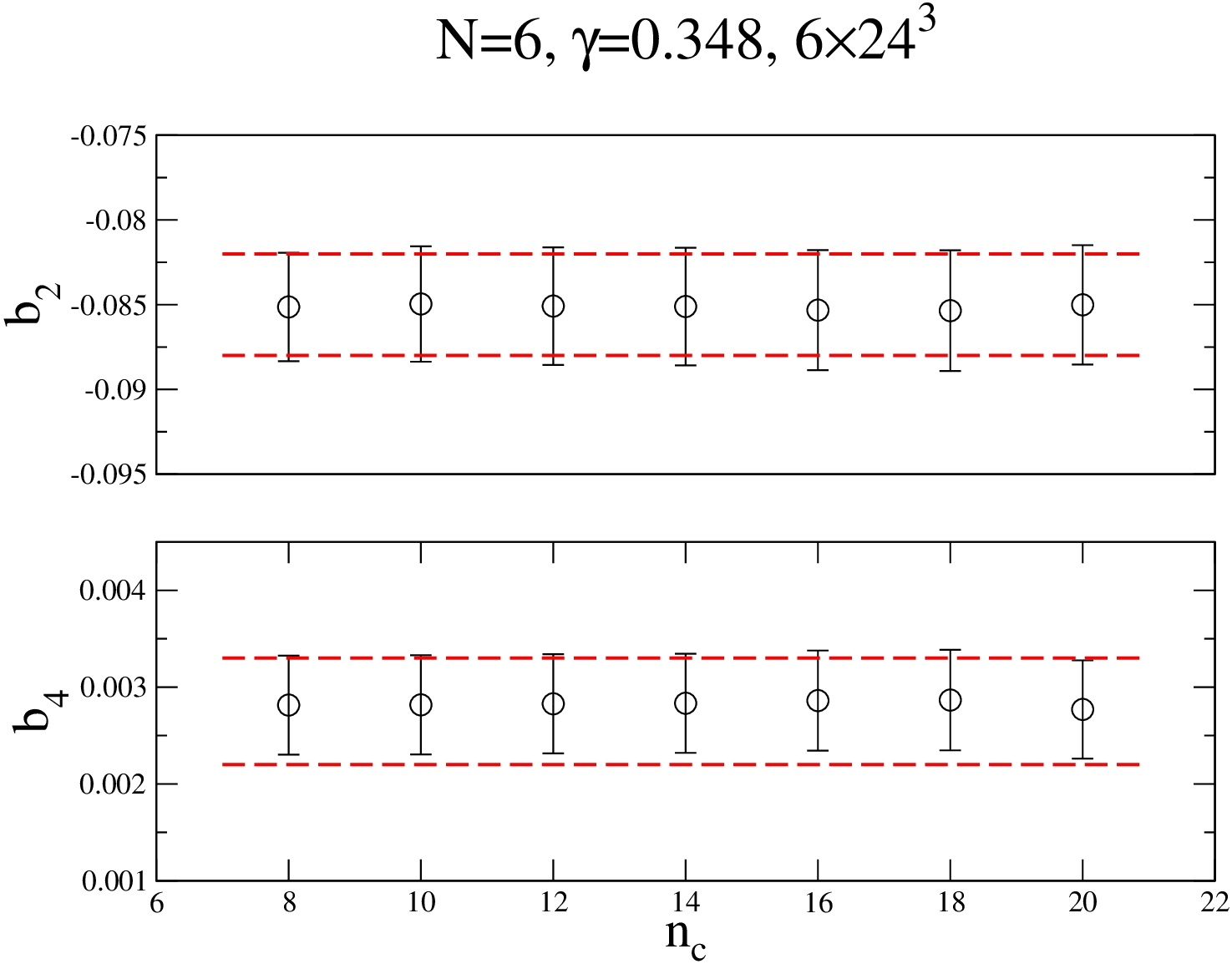}
\caption{Results for SU($6$).}
\label{test_6}
\end{figure}
}


\begin{thebibliography}{99}
\bibitem{footnoteln} This can easily be derived by recalling that the
Lagrangian is $O(N^2)$ and the large-$N$ limit is performed keeping
the coupling $g^2N$ fixed \cite{Hooft-74}.
\bibitem{n6t0} The estimates of $b_2$ and $b_4$ for $N=6$ reported in
Table~\ref{t0res} are obtained by a high-statistics MC simulation at
$\beta/(2N^2)=0.348$ on a $12^4$ lattice, averaging over $\approx
3\times 10^4$ independent measurements of $Q$ (corresponding to
$O(10^8)$ MC sweeps), which also gives $\chi_l=0.97(2)\times 10^{-4}$.
They should be taken with some caution since the continuum limit has
not been checked.
\bibitem{footnoteup}
We use a mixture of heat-bath and overrelaxation algorithms for the
link-variable updating.
\bibitem{Ntsize}
Our choice of the $L_t$ values was motivated by the fact that the cooling
method is not sufficiently stable if the temporal extent of the lattice is too small 
(see e.g. \cite{DPV-04}). 
The values of $L_t$ considered here appear safe, providing stable results. Larger
values would prevent us from obtaining the high statistics required in this study.
\bibitem{footnotestat} The MC results for $N=3$ are obtained by
averaging $O(10^5)$ independent measurements of $Q$, corresponding to
approximately $10^7-10^8$ sweeps due to the large autocorrelation time
$\tau_Q$ of $Q$~\cite{DPV-02} (e.g., $\tau_Q\approx 2\times 10^3$ for
the largest $\beta$ value, $\beta=6.437$).  In the case of $N=6$, they
are obtained from $O(10^4)$ independent measurements, which
corresponds to $O(10^7)$ sweeps (e.g., $\tau_Q\approx 400$ for the
runs at $\beta/(2N^2)=0.348,\;0.350$).
\bibitem{footnotetcres} In our analysis, and in particular to
determine the reduced temperature, we use results for the transition
points at $N=3$ taken form Ref.~\cite{LTW-04}.  For the SU($6$) gauge
theory we use $\beta_c/(2N^2)=0.34508(5)$ for $L_t=6$ (from \cite{LTW-04}),
$\sigma(\beta_c)=0.276(2)$ and
$t(L_t,\beta)=\sqrt{\frac{\sigma(\beta_c(L_t))}{\sigma(\beta)}}-1$.
\end{thebibliography}
\end{document}